\documentclass[journal]{IEEEtran}
\pdfoutput=1
\usepackage{amsmath}
\usepackage{amsfonts}
\usepackage{amsthm}
\usepackage{amssymb}
\usepackage{booktabs}
\usepackage{graphicx}
\usepackage{microtype}
\usepackage{color}
\usepackage{xcolor}

\ifCLASSOPTIONcompsoc
  \usepackage[tight,normalsize,sf,SF]{subfigure}
\else
  \usepackage[tight,footnotesize]{subfigure}
\fi

\include{definition_yilun}

\title{Multi-layer graph analysis for dynamic social networks}

\author{Brandon Oselio,~\IEEEmembership{Student Member,~IEEE,}
~Alex Kulesza,
~Alfred~O.~Hero,~III,~\IEEEmembership{Fellow,~IEEE}

\thanks{The authors are with the Department of Electrical Engineering
and Computer Science, University of Michigan, Ann Arbor, MI 48109,
USA. Tel: 1-734-763-0564. Fax: 1-734-763-8041. Emails: \{boselio, kulesza, hero\}@umich.edu.}

\thanks{This work was partially supported by ARO grant number  W911NF-12-1-0443. Parts of this paper were presented in the Proceedings of the IEEE Workshop on Computational Advances in Multi-Sensor Adaptive Processing (CAMSAP), St. Martin, Dec. 2013.
}}

\begin{document}

\maketitle

\begin{abstract}
Modern social networks frequently encompass multiple distinct types of
connectivity information; for instance, explicitly acknowledged friend
relationships might complement behavioral measures that link users
according to their actions or interests.  One way to represent these
networks is as multi-layer graphs, where each layer contains a unique
set of edges over the same underlying vertices (users).  Edges in
different layers typically have related but distinct semantics;
depending on the application multiple layers might be used to reduce
noise through averaging, to perform multifaceted analyses, or a
combination of the two.  However, it is not obvious how to extend
standard graph analysis techniques to the multi-layer setting in a
flexible way.  In this paper we develop latent variable models and
methods for mining multi-layer networks for connectivity patterns
based on noisy data.
\end{abstract}


Multi-layer networks arise naturally when there exists more than one source of connectivity information for a group of users. For instance, in a social networking context there is often knowledge of direct communication links, i.e., \textit{relational} information. Examples of relational information include the frequency with which users communicate over social media, or whether a user has sent or received emails from another user in a given time period. However, it is also possible to derive \textit{behavioral} relationships based on user actions or interests. These behavioral relationships are inferred from information that does not directly connect users, such as individual preferences or usage statistics. In this paper we show how to deal with multiple layers of a social network when performing tasks like inference, clustering, and anomaly detection.

We propose a generative hierarchical latent-variable model for multi-layer networks, and show how to perform inference on its parameters.
Using techniques from Bayesian Model Averaging \cite{raftery1995bayesian}, the layers of the network are conditionally decoupled using a latent selection variable; this makes it possible to write the posterior probability of the latent variables given the multi-layer network. The resulting mixture can be viewed as a scalarization of a multi-objective optimization problem \cite{Eh08,Ya10,NgZaEl05}. When the posterior probability functions are convex, the scalarization is both optimal and consistent with the Bayesian principle of model-averaged inference \cite{Eh08,Ge68}. 


We then step back from the Bayesian setting and discuss how multi-objective optimization can be used to perform MAP estimation of the desired latent variables. Using the concept of Pareto optimality \cite{NgZaEl05}, an entire front of solutions is defined; this allows a user to define a preference over optimization functions and tune the algorithm accordingly.
The result is a level of supervised optimization and inference that utilizes the structure of multi-layer networks without scalarization.

Experiments on a simulated example show that our method yields improved clustering performance in noisy conditions. The developed framework is then combined with the dynamic stochastic block model (DSBM) \cite{XuHe13}, which captures a variety of complex temporal network phenomena.  Finally, the multi-layer DSBM is applied to a real-world data set drawn from the ENRON email corpus. This example illustrates how we can combine two layers of a network to explore complex connections through both time and layer mixing parameters.

\section{Multi-layer networks}

A multi-layer graph $G = (\mathcal V, \mathcal E)$ comprises vertices $\mathcal V = \{ v_1, \ldots, v_p \}$, common to all layers, and edges $\mathcal E = (\mathcal E_1, \ldots, \mathcal E_L)$ on $L$ layers, where $\mathcal E_i$ is the edge set for layer $i$.

\begin{figure}
\centering
\includegraphics[width = 2.4in]{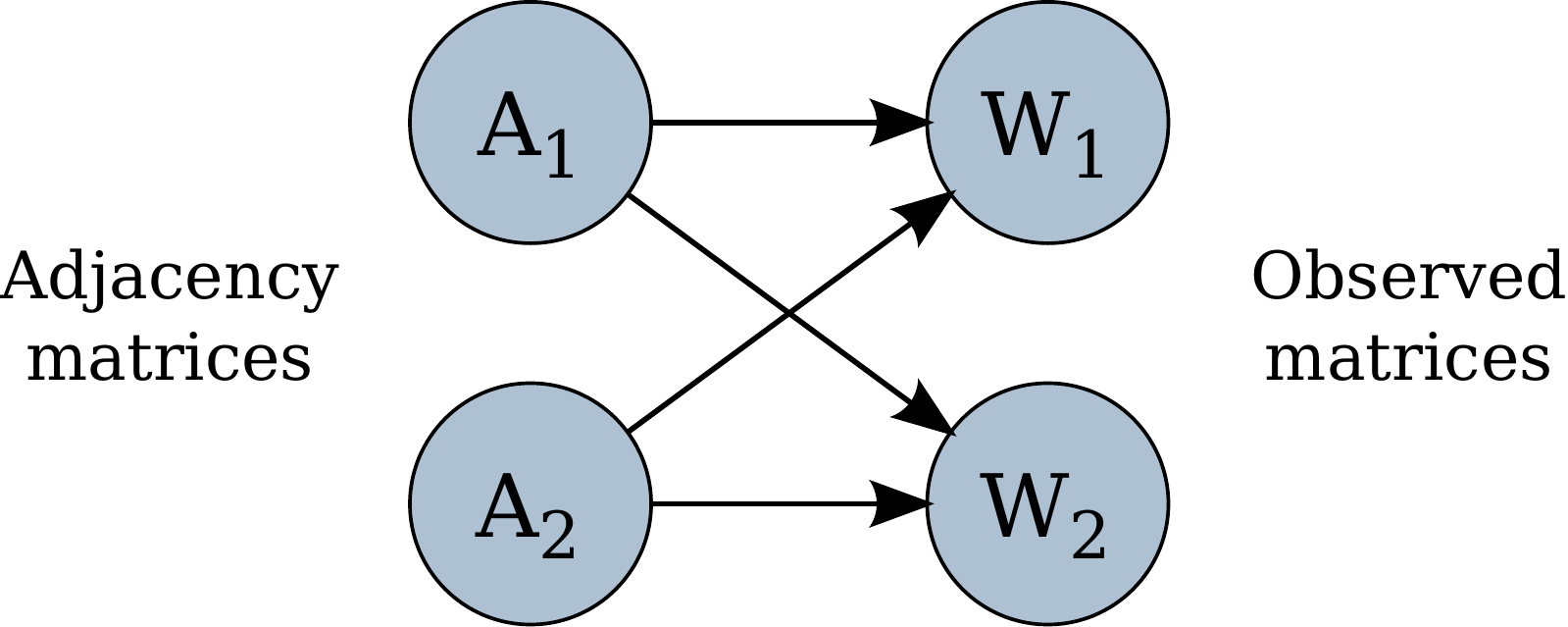}
\caption{Adjacency and Observation Matrices. This graphical model depicts how the latent adjacency matrices can affect the observervation matrices. Note that the observation matrices are dependent on all adjacency matrices in general.}
\label{Model1}
\end{figure}

In the real-world network setting, we will assume that the observed data are noisy reflections of a true underlying multi-layer graph.  For convenience we will work with adjacency representations, letting $A_i \in \mathbb{R}^{p \times p}$ be the true adjacency matrix of layer $i$, and $W_i \in \mathbb{R}^{p \times p}$ the corresponding observed adjacency matrix.  Figure~\ref{Model1} depicts the model graphically.

In some cases $W_i$ might be binary, reflecting merely the presence or absence of a connection---for instance, whether two users were seen to communicate.  In other settings, such as measuring temporal or content correlation scores between users, the entries of $W_i$ could be real-valued.  The goal is to estimate $A_1, \ldots, A_L$ given the observations $W_1, \ldots, W_L$.  Using standard parametric methods this will require computing the posterior distribution of $A_1, \ldots, A_L$, which can be difficult given the number of parameters.  Specifically, the influence of $A_1, \ldots, A_L$ on a single $W_i$ is difficult to measure, as the dependencies are unspecified. 

\section{Hierarchical model description}

A hierarchical model is proposed that simplifies this inference procedure by conditionally decoupling $W_1, \ldots, W_L$. For simplicity, we specialize to the case where $L = 2$. This also allows us to view the networks in the setting described in the introduction: one layer of the network represents the observed extrinsic relationships between users, and the other layer represents their correlated intrinsic behaviors.

\begin{figure}
\centering
\includegraphics[width = 3.0in]{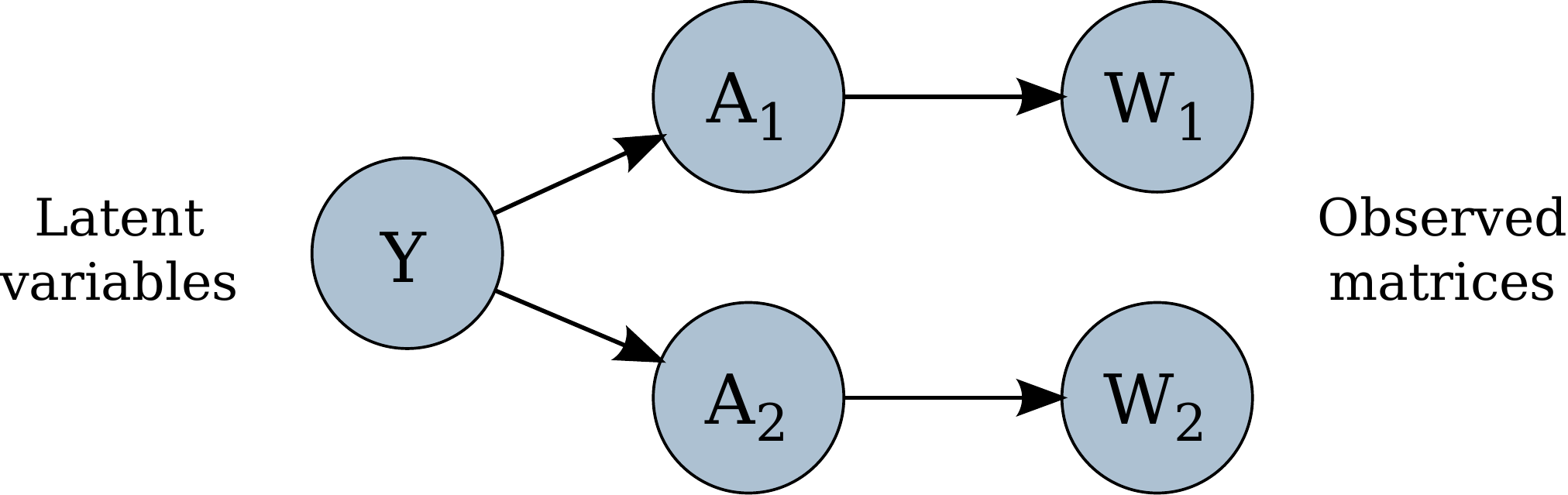}
\caption{General Latent Variable Model. This model represents a latent variable model, in which a set of variables $Y$ control the distributions of the adjacency matrices and through them the observation matrices.}
\label{Model2}
\end{figure}

We introduce a latent variable denoted $Y$ (see Figure~\ref{Model2}) that conditionally decouples the posterior distributions of the two layers:
\begin{align}
  P(W_1, W_2|A_1,A_2, Y) &= P(W_1|A_1,Y) P(W_2|A_2,Y) \\
  P(W_1,W_2|A_1,A_2) &= \nonumber\\
  \int P(W_1, W_2&|A_1,A_2, Y) P(Y|A_1, A_2)dY~.
\end{align}

Shifting the focus from the adjacency matrices $A_1, A_2$, to the latent variable $Y$, using $Y$ as a compact description of how these adjacencies combine to form the multi-layer network structure. It is possible to write down the posterior distribution for $Y$ as
\begin{equation}
  P(Y|W_1,W_2) = \sum_{A_1,A_2} P(Y|A_1,A_2) P(A_1,A_2|W_1,W_2)~.\nonumber
\end{equation}

\section{Posterior mixture modeling}
\label{sec:posterior}

\begin{figure}
\centering
\includegraphics[width = 2.5in]{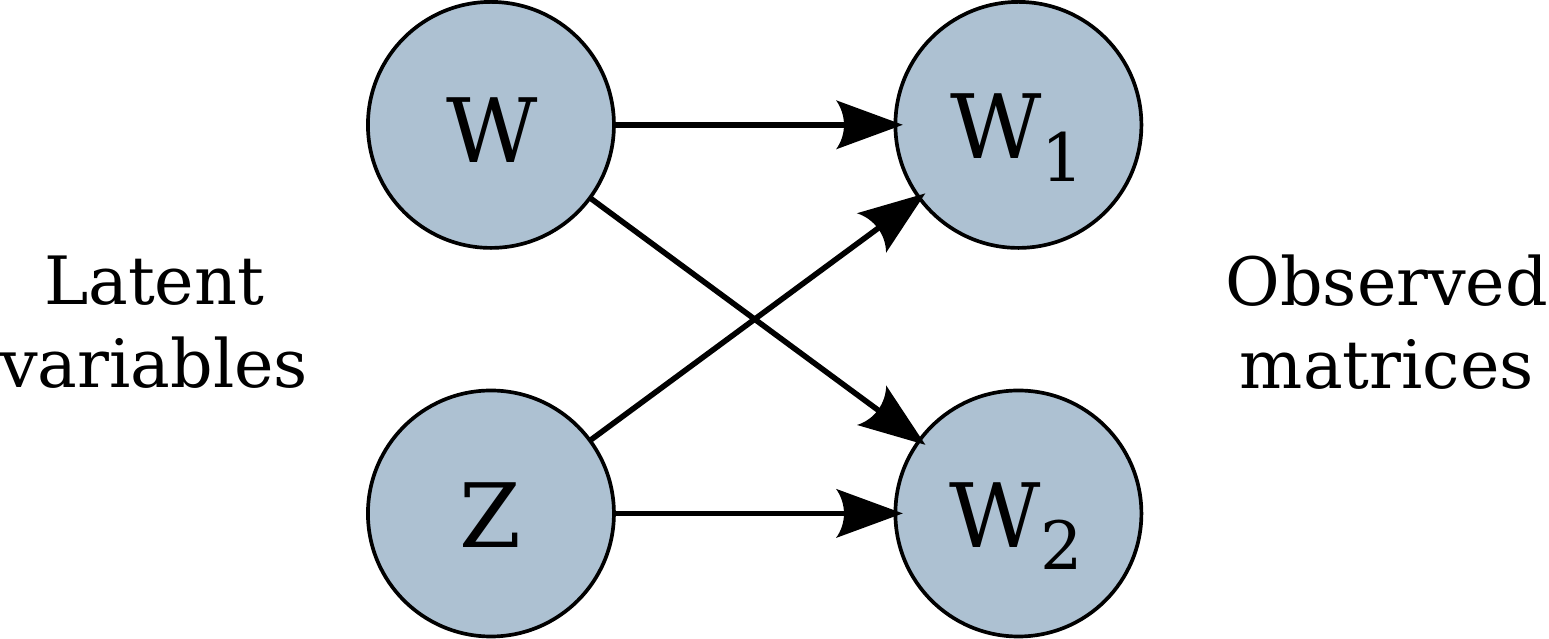}
\caption{Model with Similarity Matrix and Selection Variable. We introduce the similarity matrix $W$	 and the selection variable $Z$ to describe our latent variable model. Conditioning on W and Z, we assume that the two layers are independent from each other.}
\label{Model3}
\end{figure}

Consider the graphical model shown in Figure~\ref{Model3}. We have collapsed the $A_1, A_2$ variables with the observed data $W_1, W_2$, because we are mainly interested in inferring $W$, and $W_i$ can be considered a representation of the real connectivity. 

Following the previous model, we have decomposed $Y = (W, Z)$, where $W \in \mathbb{R}^{p \times p}$ is a latent adjacency or similarity matrix describing the underlying connections between vertices, and $Z \in \{1, 2\}$ is a model selection variable, $P(Z = 1) = \alpha$, and $P(Z = 2) = 1 - \alpha$.  Here there is the implicit assumption  a common connectivity structure $W$ informs all layers of the network. In a sense, the model produces observed matrices that correspond to multiple views of the latent variable $W$.  The model selection variable $Z$ will decouple the posterior distribution of $W$ given both layers into a weighted sum of marginalized posteriors given each individual layer.  

The prior for $W$ is $P(W)$, left unspecified for now. The
distributions $P(W_1|W,Z)$ and $P(W_2|W,Z)$ are in general
task-dependent (e.g., they could be Gaussian, Wishart, Bernoulli,
etc.), but we will make the simplifying assumption that $Z$ acts as a
selector variable, so that $W$ and $W_1$ are conditionally independent
given $Z=2$, and likewise $W$ and $W_2$ are conditionally independent
when $Z=1$.  Formally, using the notation $P_z$ to denote conditioning
on $Z=z$, we have
\begin{align}
  P_2(W_1 | W) &= P_2(W_1)\\
  P_1(W_2 | W) &= P_1(W_2)~.
\end{align}

We are interested in the posterior distribution of the latent variable
$W$ given the observed variables $W_1, W_2$:
\begin{align}
  P(W|&W_1,W_2) \\
  &= P(W,Z=1|W_1,W_2) + P(W,Z=2|W_1,W_2)\\
  &= P(W|W_1,W_2,Z=1)P(Z=1|W_1,W_2) \nonumber\\
  &\quad\quad+ P(W|W_1,W_2,Z=2)P(Z=2|W_1,W_2)\\
  &= \xi P(W|W_1,W_2,Z=1) \nonumber\\
  &\quad\quad+ (1-\xi)P(W|W_1,W_2,Z=2)~,
\end{align}
where $\xi = P(Z=1|W_1,W_2)$.  Let's consider the first term.  We have
\begin{align}
  P(W|W_1,W_2,Z=1) &= \frac{P(W,W_1,W_2,Z=1)}{\sum_{\hat W} P(\hat W,W_1,W_2,Z=1)} \\
  &= \frac{P(W)P_1(W_1|W)P_1(W_2)}{\sum_{\hat W} P(\hat W)P_1(W_1|\hat W)P_1(W_2)}~.
 \label{eq:bayes}
\end{align}
Since $P_1(W_2)$ does not depend on $W$, it factors out of the sum in
the denominator and cancels; thus (\ref{eq:bayes}) becomes
\begin{align}
  P(W|W_1,W_2,Z=1) &= \frac{P(W)P_1(W_1|W)}{P_1(W_1)}~.
\end{align}
Performing the same computation on the other side and combining, we have
\begin{align}
  P(W|&W_1,W_2)\\
  &=\xi \frac{P(W)P_1(W_1|W)}{P_1(W_1)}
  + (1-\xi) \frac{P(W)P_2(W_2|W)}{P_2(W_2)}\\
  &= P(W) \left[\gamma_1 P_1(W_1|W) + \gamma_2 P_2(W_2|W)\right]~,
  \label{eq:firstobj}
\end{align}
where $\gamma_1 = \xi / P_1(W_1)$ and $\gamma_2 = (1-\xi) /
P_2(W_2)$ are constants with respect to $W$.  If we assume the prior
on $W$ is uniform, then the MAP estimate of $W$ is also the maximum
likelihood estimate, which can be written as
\begin{align}
  \mathrm{argmax}_W \left[\gamma_1 P_1(W_1|W) + \gamma_2 P_2(W_2|W)\right]~.
  \label{eq:secondobj}
\end{align}

The above solutions describe not just one MAP estimate of $W$, but rather a family of MAP estimates, based on the priors that we implicitly assign to each model by choosing a specific value of $\alpha$ (which affects $\xi$ and $\gamma$ in turn). Qualitatively, this can be viewed as determining a relative confidence parameter between the networks; if $W_1$ is more trusted than $W_2$, then $\alpha$ would be greater than 0.5.

As an example, assume that both $P(W_1|W)$ and $P(W_2|W)$ are isotropic Gaussians, i.e.,
\begin{align}
  P(W_1|W)& = \mathcal{N}(W, \sigma_1^2 I_p) \\
  P(W_2|W)& = \mathcal{N}(W, \sigma_2^2 I_p)~.
\end{align}
Then the solution for $\hat{W}$ has the form
\begin{equation}
  \hat{W} = \beta W_1 + (1 - \beta) W_2~,
  \label{eq:scalarization}
\end{equation}
for some choice of $0 \leq \beta \leq 1$.

A proof of this is given in Appendix A. In the non-isotropic, non-Gaussian case the solution will not have such a simple form. However, numerical methods can be used to compute the solution (\ref{eq:secondobj}).



\section{Simulation Example}

\begin{figure}
\centering
\includegraphics[width = 3in]{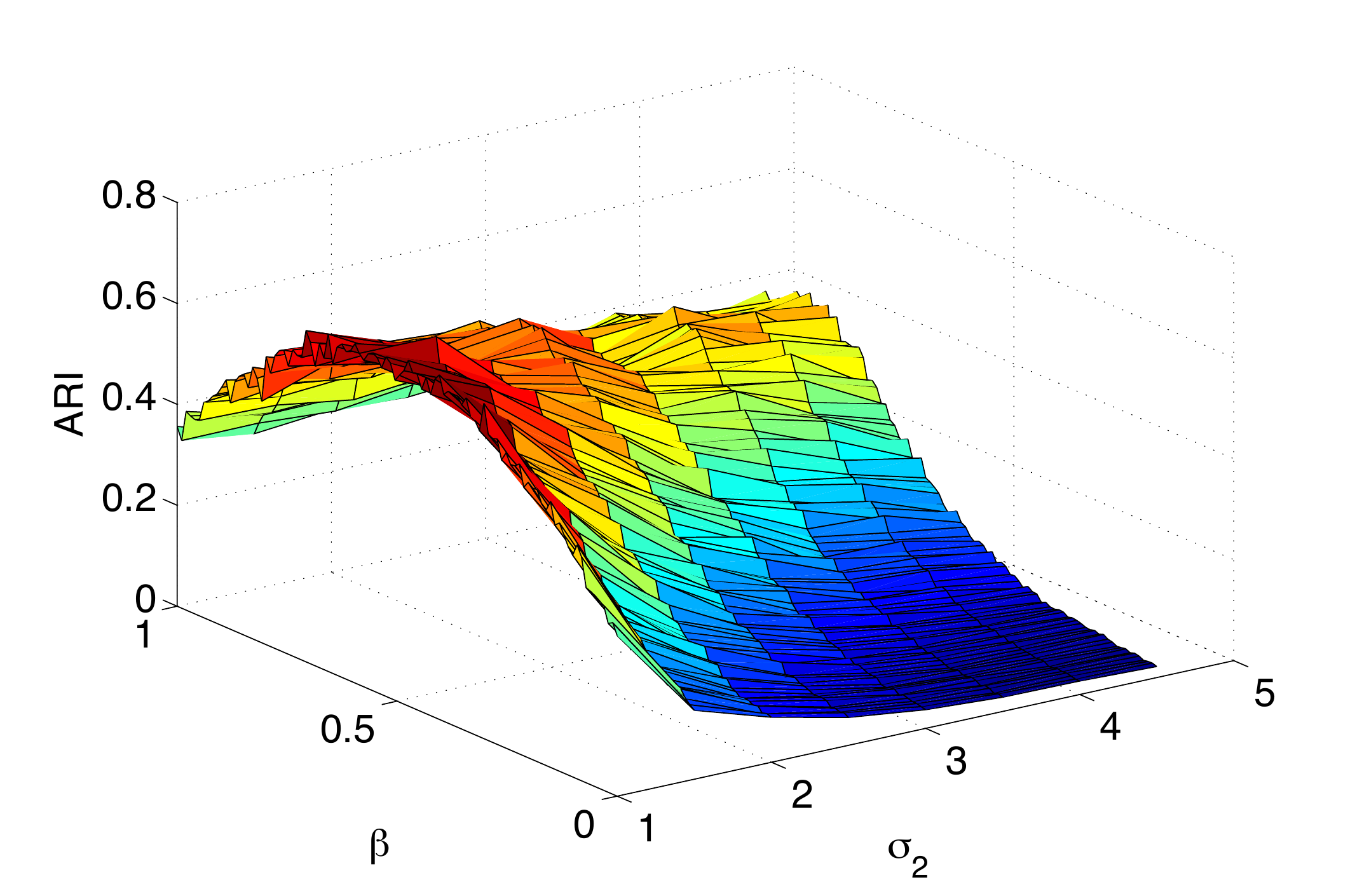}
\caption{Clustering Simulation. This surface plot shows the ARI for different simulations of $\sigma_2$ and $\beta$. Note that for all levels of $\sigma_2$, a $\beta$ that is around 0.5 tends to produce the best clustering.}
\label{Clus1}
\end{figure}

We use simulations to show that clustering of nodes in a weighted graph can be improved using the MAP estimate of $W$. This simulation example uses the Bayesian posterior representation, where $P(W_1|W)$ and $P(W_2|W)$ are isotropic multivariate Gaussian distributions with (posterior) mean $W$. Two weighted random graphs with 500 nodes are constructed with 10 known clusters of equal size.  The weights between nodes in the same cluster are independently generated from the normal distribution $\mathcal{N}(5, 0.5)$, and the edge weights between nodes that are not in the same cluster are independently generated from the normal distribution $\mathcal{N}(4.7, 0.5)$. The dichotomy between these edge weights is to simulate the underlying community structure with variability. The networks are then corrupted with i.i.d. Gaussian noise on each edge weight with zero mean and different variances. Specifically, the first network layer is corrupted with additive noise distributed as $\mathcal{N}(0, \sigma_1)$ and the second layer is corrupted with additive noise distributed as $\mathcal{N}(0, \sigma_2)$. This setup corresponds to the form of $\hat W$ that is derived in (\ref{eq:scalarization}). For various choices of mixing parameters $\beta$, the combined network $\hat W$ is calculated and then clustered using a spectral clustering algorithm \cite{Lu07}. The spectral clustering algorithm finds the eigenvectors of the graph Laplacian $L = D - A$, where $D, A$ are the degree and adjacency matrix obtained from $\hat W$. The Adjusted Rand Indices (ARI) \cite{HuAR85}  are computed in comparison to the true clustering structure; this gives us a measure of the quality of the clustering. For each of several different levels of noise variance, this experiment is run 50 times, and the results are averaged. Figure~\ref{Clus1} computes the solution~(\ref{eq:secondobj}), and shows that using (\ref{eq:firstobj}) to estimate the mixture of networks improves clustering when compared to using only one layer of the network, as expected.

\begin{table}
\caption{Variances and ARI Scores}
\label{ARI} \centering
\begin{tabular}{ cccc }
\toprule
$\sigma_1$ & $\sigma_2$ & Max ARI & $\beta$ \\ \midrule
        1         &          1        &  0.6843  &      0.4747         \\
        1         &          1.5     &  0.6561  &      0.5859         \\
        1         &          2        &  0.5564  &      0.6364         \\
         1         &          2.5     &  0.5649  &      0.6970         \\
         1         &          3        &  0.4918  &      0.7879         \\
         1         &          3.5     &  0.5209  &      0.7475         \\
         1         &          4        &  0.4809  &      0.7374         \\
         1         &          4.5     &  0.4653  &      0.7879       \\\bottomrule
\end{tabular}
\end{table}

\section{Pareto summarizations}
Of course, in practice it may be difficult to effectively set the prior parameter $\alpha$. In such cases we can generate a family of MAP estimates and apply multiple-objective ranking techniques.  In particular, one can view the maximization (\ref{eq:secondobj}) of the combined posterior distributions as a particular scalarization of a multi-objective optimization problem. However, there are other solutions to multiple objective optimization that do not use linear scalarization, such as Pareto front analysis \cite{jin2006multi, hero2004multicriteria, NIPS2012_0395}.

Consider the multi-objective optimization problem
\begin{equation}
  \hat{W} = \mathrm{argmin}_W [f_1(W), f_2(W)]~,
  \label{eq:multiobj}
\end{equation}
where the minimization in (\ref{eq:multiobj}) is in the sense of multi-objective minimization, to be made clear below. For the model derived in Section~\ref{sec:posterior}, we have $f_1(W) = -P_1(W|W_1)$ and $f_2(W) = -P_2(W|W_2)$, with (\ref{eq:multiobj}) being interpreted in terms of linear scalarization using weighting coefficients $\gamma$ and $1-\gamma$:

\begin{align}
\hat{W} = \mathrm{argmin}_W \left[ \gamma f_1(W) + (1 - \gamma) f_2(W) \right]~.
\end{align} 

An alternative to the scalarization approach is a ranking approach that seeks to find a family of solutions $W$ that would be highly ranked by any scalarization, linear or non-linear. This leads to the idea of Pareto optimization. A solution to a multi-objective optimization problem is said to be weakly Pareto optimal (or weakly non-dominated) if it is not possible to improve any single objective function without lowering some other objective function \cite{Eh08}, \cite{Ya10}. More formally, we say that a solution $W_1$ dominates a solution $W_2$ if $f_i(W_1) \le f_i(W_2)$ for every objective function $f_i$ and there exists some $j$ such that $f_j(W_1) < f_j(W_2)$. The first Pareto front is the set of weakly non-dominated points.

\begin{figure}
\centering
\includegraphics[width = 2.3in]{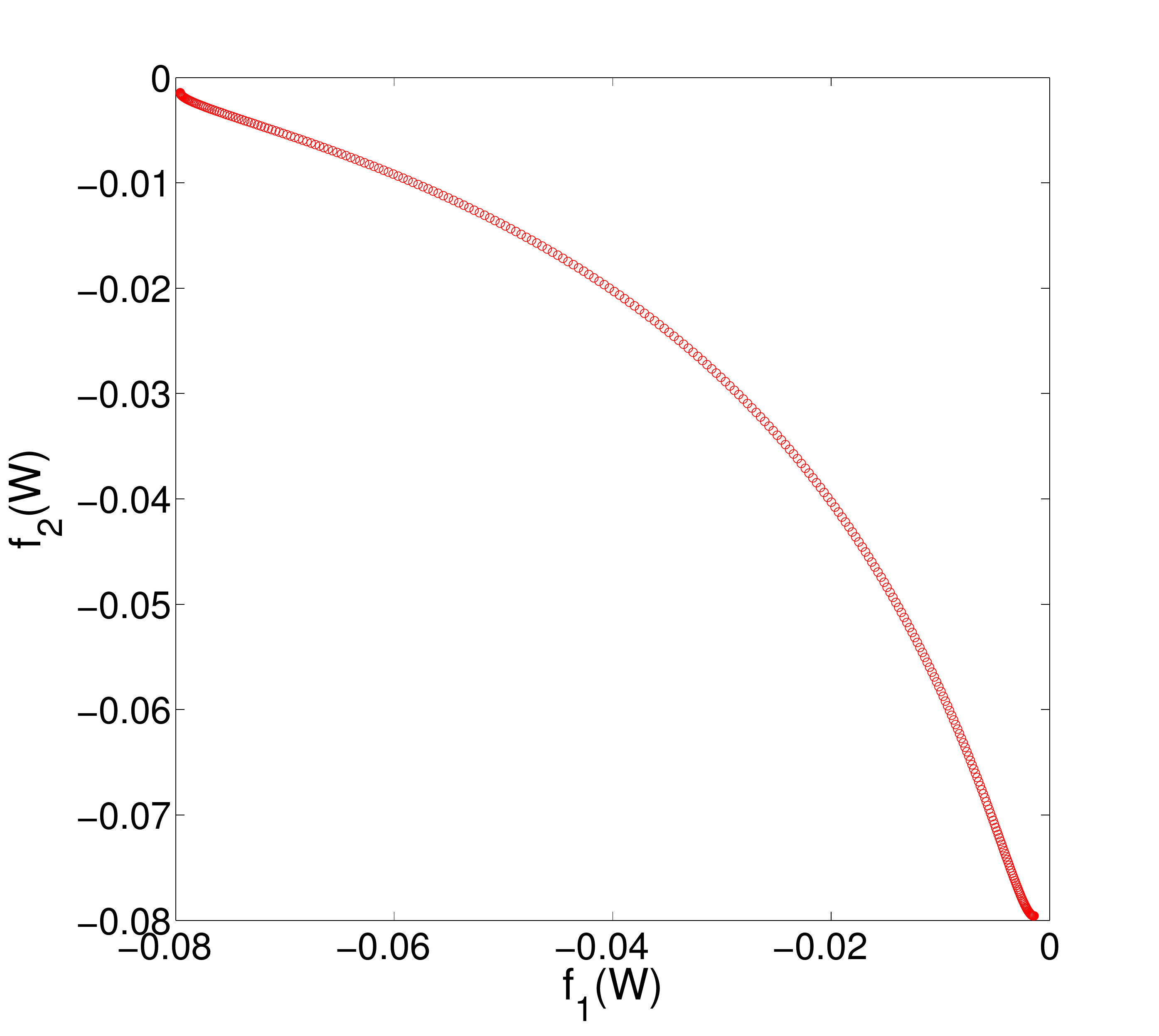}
\caption{Pareto front for two Gaussians.  A convex Pareto front would bulge toward the lower left corner, but this plot demonstrates that even relatively simple objective equations can have extremely non-convex Pareto fronts.}
\label{Pareto}
\end{figure}

In terms of finding Pareto optimal points, the linear scalarization technique discussed above can identify the complete Pareto front when the solution space is a convex set and the individual objective functions are convex functions on the solution space \cite{Ge68}. However, if these convexity conditions are not met, the scalarization technique will not find the entire Pareto front. Often, the posterior distributions in (\ref{eq:multiobj}) are not convex. Figure~\ref{Pareto} shows an example of the Pareto front of a multiobjective optimization, where $f_1$ and $f_2$ are the two dimensional pdfs of normal distributions, as shown below:
\begin{align}
f_i(W) &= \left(2\pi \right) ^{-n/2} | \Sigma_i |^{-\frac{1}{2}}e^{-\frac{1}{2}(W - W_i)^T  \Sigma_i^{-1}(W - W_i)} \\
W_1 &= 
\begin{bmatrix}
    10\\
    8
  \end{bmatrix}, 
W_2 = 
\begin{bmatrix}
    8\\
    10
  \end{bmatrix}, 
\Sigma_1 =  \Sigma_2 = 2I_2~.
\end{align}
Even this relatively simple distribution has a non-convex Pareto front; note that minimizing a linear combination of $f_1$ and $f_2$ can only find optima at the extremes of the curve, and does not explore the interior, which may be more useful for some applications.

This example motivates further research into generating MAP estimates in this manner, as finding the Pareto front could give us an advantage when attempting to infer parameters of the model as we do above, or perform some other common task; see for instance \cite{OsKuHe14}.

\section{Stochastic Block Models and the DSBM}
\label{sec:DSBM}

Consider a single layer network. Often we are interested in networks that are expected to have some community structure. A community is defined as a subset of nodes that behave similarly to each other, where similarity is determined according to some fixed criterion. This allows for a more interesting community structure than just using the density of connections in a group, i.e., creating communities based on high intra-connectivity between nodes. For instance, one group may exhibit strong interconnection with another group, but only moderate connectivity within themselves. A Stochastic Block Model (SBM) is one way to model such community structure. \cite{WaWo87, XuKlHe14}.

Consider a network with $N$ nodes that we expect to fall in $K$ classes, where $c \in \mathbb{R}^N$ is a known class membership vector. In this setup we are considering binary relationships between nodes, and so a connectivity matrix $A = \{a_{xy}\} \in \mathbb{R}^{N \times N}$ is observed. The parameters for a standard SBM are prior probabilities of edges occurring between nodes within and across classes. Specifically, let $\Theta$ be a matrix of class probabilities called the Bernoulli parameter matrix, where $\theta_{ij}$ is the probability of a link forming between a node in class i and class j. While the graph adjacency matrix will be $ N \times N$, $\Theta \in \mathbb{R}^{K \times K}$ and is symmetric. Letting $S_i = \{ x\ |\ c(x) = i \}$, it can be shown (\cite{XuHe13}) that the MLE of $\theta_{ij}$ is

\begin{align}
\hat{\theta}_{ij} &= \frac{m_{ij}}{n_{ij}} \\ 
m_{ij} &= \sum_{x \in S_i} \sum_{y \in S_j} a_{xy} \\
n_{ij} &= 
\begin{cases}
|S_i||S_j|, &, i \neq j \\
|S_i|(|S_i| - 1), &, i = j
\end{cases}~.
\end{align}

This estimate of $\Theta$ (which we call $Y$) can be used to explore the structure of the network. When the class membership vector $c$ is unknown, the SBM can be modified to simultaneously estimate $c$ and the Bernoulli matrix $\Theta$ \cite{YaChZh11}.

The SBM accounts for community structure, but does not account for temporal changes in the network. One solution to this problem would be to fit a SBM to every time step in the sequence. This approach, however, fails to take advantage of information from previous time steps, and it does not encourage the class membership to evolve smoothly over time. Recently, the Dynamic SBM (DSBM) has been introduced to account for some of these effects  \cite{YaChZh11, HoSoXi11, XuHe13}. 

The DSBM of \cite{XuHe13} employs an extended Kalman filter (EKF) to track temporal changes in the network. Two types of DSBM were introduced in \cite{XuHe13}: one that is given the class membership \textit{a priori}, and another that estimates the class memberships along with the other SBM parameters. For the benefit of the reader, the {\em a priori} DSBM is briefly reviewed below. The DBSM is based on the following simple linear model observation:

\begin{equation}
Y^t = \Theta^t + z^t~,
\label{lin_model}
\end{equation}
where $z^t$ is i.i.d. zero-mean Gaussian noise and $\Theta^t$ is an unknown matrix of  Bernoulli parameters at time $t$. Because the elements of $\Theta^t$ must be between 0 and 1, the DSBM uses a logistic transform to map them onto the real line:

\begin{equation}
  \psi_{ij} = \log(\theta_{ij}) - \log(1 - \theta_{ij}) \in (-\infty, + \infty)~.
\end{equation}

Since the logistic transform is invertible, (\ref{lin_model}) can be written as

\begin{equation}
Y^t = h(\psi^t) + z^t~.
\label{lin_model2}
\end{equation}

A linear state space model for the time evolution of the logistically transformed parameters $\psi^t$ is assumed. With this state space model for $\psi^t$ and the observation model (\ref{lin_model2}), an extended Kalman filter estimator can be implemented to produce state estimates $\hat{\psi}^{t|t-1}$ from which the SBM parameters can be tracked over time:
\begin{align}
\hat{\psi}^{t|t-1} &= F^t\hat{\psi}^{t-1|t-2} + K^{t|t-1} \eta^t~,
\end{align}
where  $\eta^t =Y^t - H^t \hat{\psi}^{t|t-1}$ is the Kalman innovation process, $K^{t|t-1}$ is the ($\hat{\psi}^{t|t-1}$-dependent) Kalman gain, and $H^t$ is the Jacobian of $h(\hat{\psi}^{t|t-1})$.
Once the inference is complete, the Kalman estimate is then mapped back into Bernoulli parameters.  When the class memberships are unknown, the DSBM can be modified to estimate these memberships and the probability parameters simultaneously [7].    For the ENRON data experiment described below, we implemented a multi-layer extension of the {\em a priori} DSBM in [7] using a simple random walk state space model ($F^t=I$, the identity matrix).

\section{ENRON Example}
The proposed dynamic SBM multi-layer community detection approach of Section \ref{sec:DSBM} is illustrated on real-world ENRON email data set\footnote{\url{http://www.cs.cmu.edu/~enron}}. This data set consists of approximately a half million email messages sent or received by 150 senior employees of the ENRON Corporation. These emails were made publicly available as a result of the SEC investigation of the company in 2002, and constitute one of the largest publicly available email repositories. This dataset represents a unique opportunity to examine private email messages in a corporate setting. This is rare due to privacy concerns and proprietary information, but the ENRON dataset is for the most part untouched, except for a few emails that were specifically requested to be removed. In addition to the raw emails, the dataset also contains the job title of the employees that are included. This is useful to separate the employees into classes, so that we may examine their behavior using the DSBM and its related techniques.

To explore the multi-level structure, two layers are extracted from the ENRON dataset. As discussed previously, one layer represents the extrinsic, "relational" information between users, and the other represents intrinsic, "behavioral" information between users. The network layers are extracted from the data as follows. First, a \textit{relational} network is recovered from the headers of emails by identifying the sender and receiver(s) of each message, including Cc and Bcc recipients. For each week in the dataset, a separate network of employees is constructed from the emails sent during that week.  

A second set of \textit{behavioral} networks are recovered using the contents of email messages. On the same weekly basis the contents of all emails originating from each user are combined to form long ``documents''. Only emails that are sent by the user are considered, which is different from the relational case. This is to obtain a better representation of each user's individual writing habits, as opposed to the writing habits of them and their peers. These emails combine to produce a dictionary of words from which term frequency-inverse document frequency (TF-IDF) scores are calculated  \cite{BaCaCa11}.  TF-IDF scores are commonly used for identifying important words in text analysis, and are computed using
\begin{align}
  \text{tf}(t,d) &= \frac{f(t,d)}{\max_{\hat t} f(\hat t,d)} \\
  \text{idf}(t) &= \log\left( \frac{|D|}{N(t,D)} \right) \\
  \text{score}(t, d) &= \text{tf}(t,d)\text{idf}(t)~,
\end{align}
where $f(t,d)$ is the frequency of term $t$ in document $d$, $N(t, D)$ is the number of documents in which the term $t$ appears, and $|D|$ is the size of the document corpus, which in this case is the number of active network nodes. For each active user (document), a TF-IDF score is computed for each word in the dictionary. 

\begin{figure*}[!t]
\centering
  \subfigure[Relational DSBM Parameters]{\includegraphics[width=\textwidth]{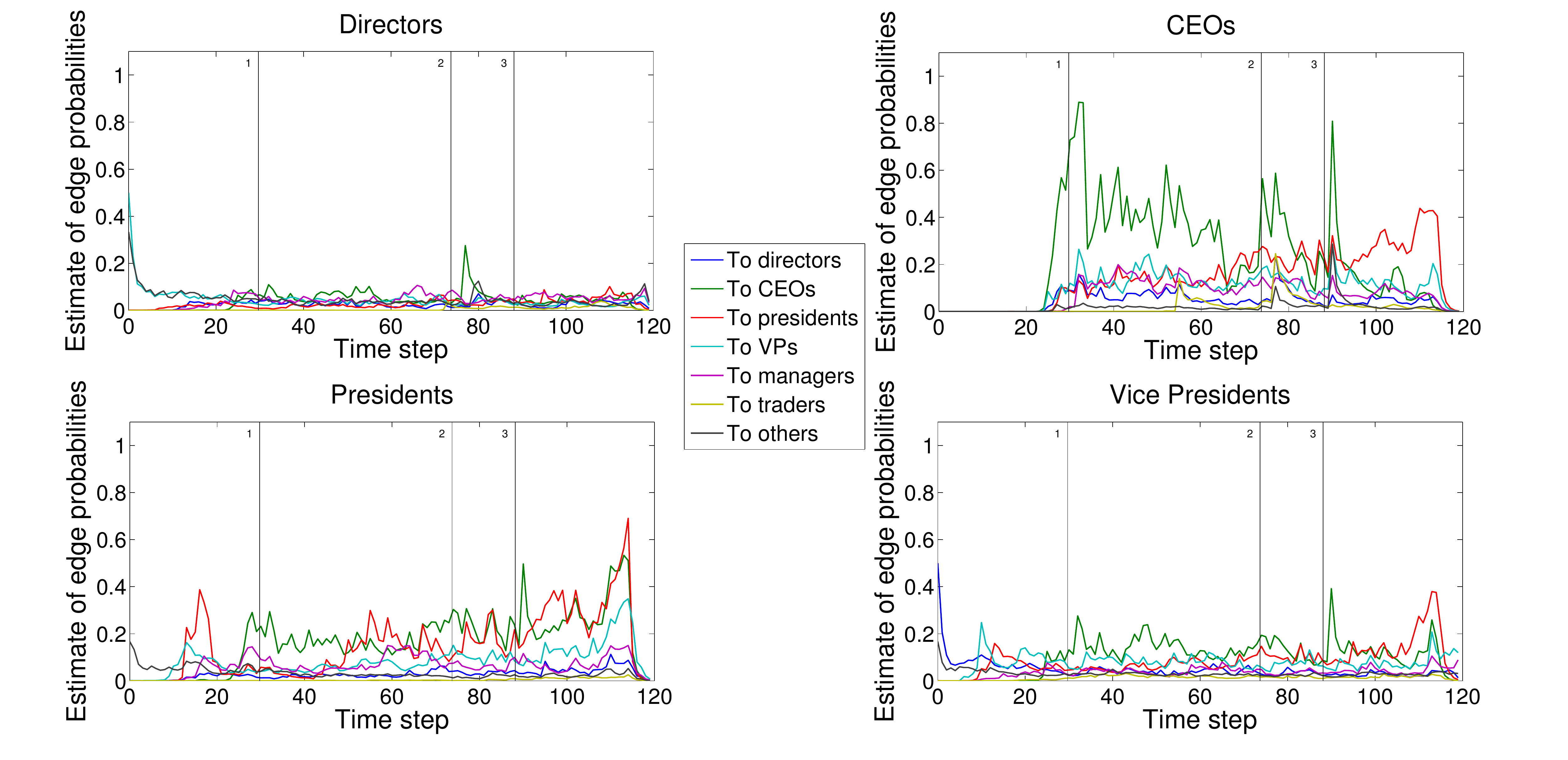}
    \label{rel_DSBM}}\\
  \subfigure[Behavioral DSBM Parameters]{\includegraphics[width=\textwidth]{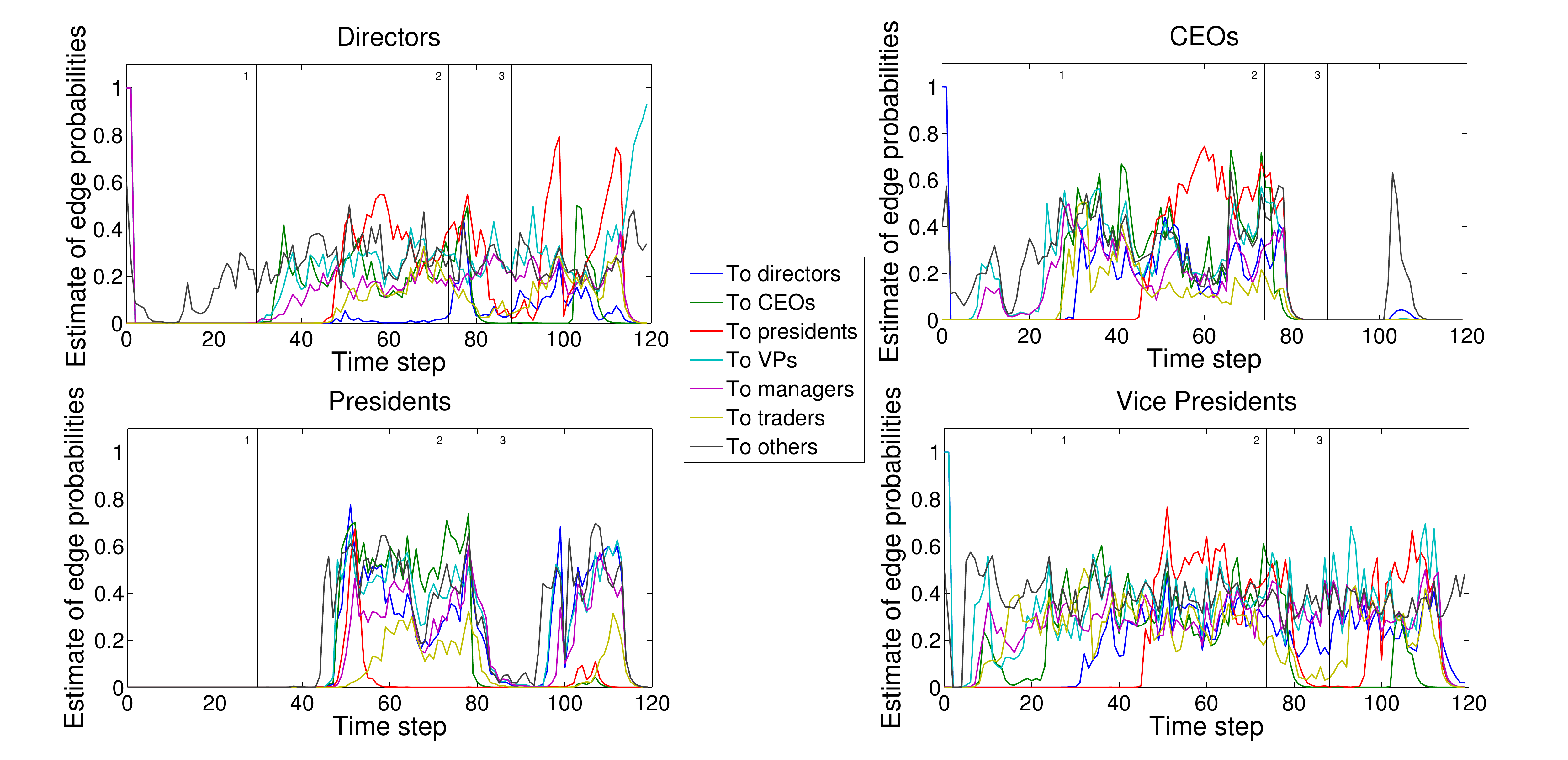}
    \label{beh_DSBM}}
\caption{DSBM Simulation Results. These graphs show the estimated DSBM parameters for different classes, and how they evolve over time. (a) is the evolution of the DSBM parameters from the relational layer, while (b) is the evolution of parameters from the behavioral layer. }
\label{DSBM_params}
\end{figure*}

Using the vector of TF-IDF scores for each user, we measure the cosine similarity of each user by taking dot products in order to obtain a similarity matrix $W$. Again, this is done for every week in the relevant time period, creating a second dynamic network with weighted edges. However, since we started in the SBM framework, it is necessary to transform the weighted edge network into a binary network.  To do this, the similarity scores are thresholded. To be roughly consistent with the density of the relational network, we keep the top 15\% greatest correlations between users at each time step, setting all other connections to 0. This allows us to create networks of similar sparsity level.

The above procedure yields a two-layer binary dynamic network that we can use to obtain insight into the structural dynamics of the ENRON data. To do so, we extended the dynamic stochastic block model (DSBM) \cite{XuHe13,HoLaLe83} to the multi-layer setting. We group employees by their role in the company (CEO, President, Director, etc.). Thus, the DSBM class memberships are known \textit{a priori}, and the {\em a priori} DSBM described in Section \ref{sec:DSBM} can be implemented to estimate the Bernoulli parameters, which predict the likelihood of an edge between users from any pair of groups. 

Figure~\ref{rel_DSBM} and Figure~\ref{beh_DSBM} shows some of the estimated Bernoulli parameters for different classes when the DSBM is run on the two layers separately. Figure~\ref{rel_DSBM} represents the evolution of the relational layer, while the Figure~\ref{beh_DSBM} represents the behavioral layer. The DSBM was run over a 120 week period, from December 6th, 1999 to March 27th, 2002. The vertical lines represent important events in the ENRON time line. Line 1 corresponds to ENRON releasing a code of ethics policy. It is also the first time that the company's stock reached above \$90. Line 2 corresponds to their stock closing below \$60. This was a critical point in the timeline, because the company began losing many partnerships, including one to create a video-on-demand system. In this same month, a few of the employees had begun to communicate the uneasiness with ENRON's accounting practices. Line 3 is the week of Jeffrey Skilling's resignation. A mere month after his resignation as CEO, the SEC began their official inquiry into ENRON.  These events are chosen as a baseline to compare the two layers of the network.

\begin{figure*}[h!t]
\centering
\includegraphics[width=\textwidth]{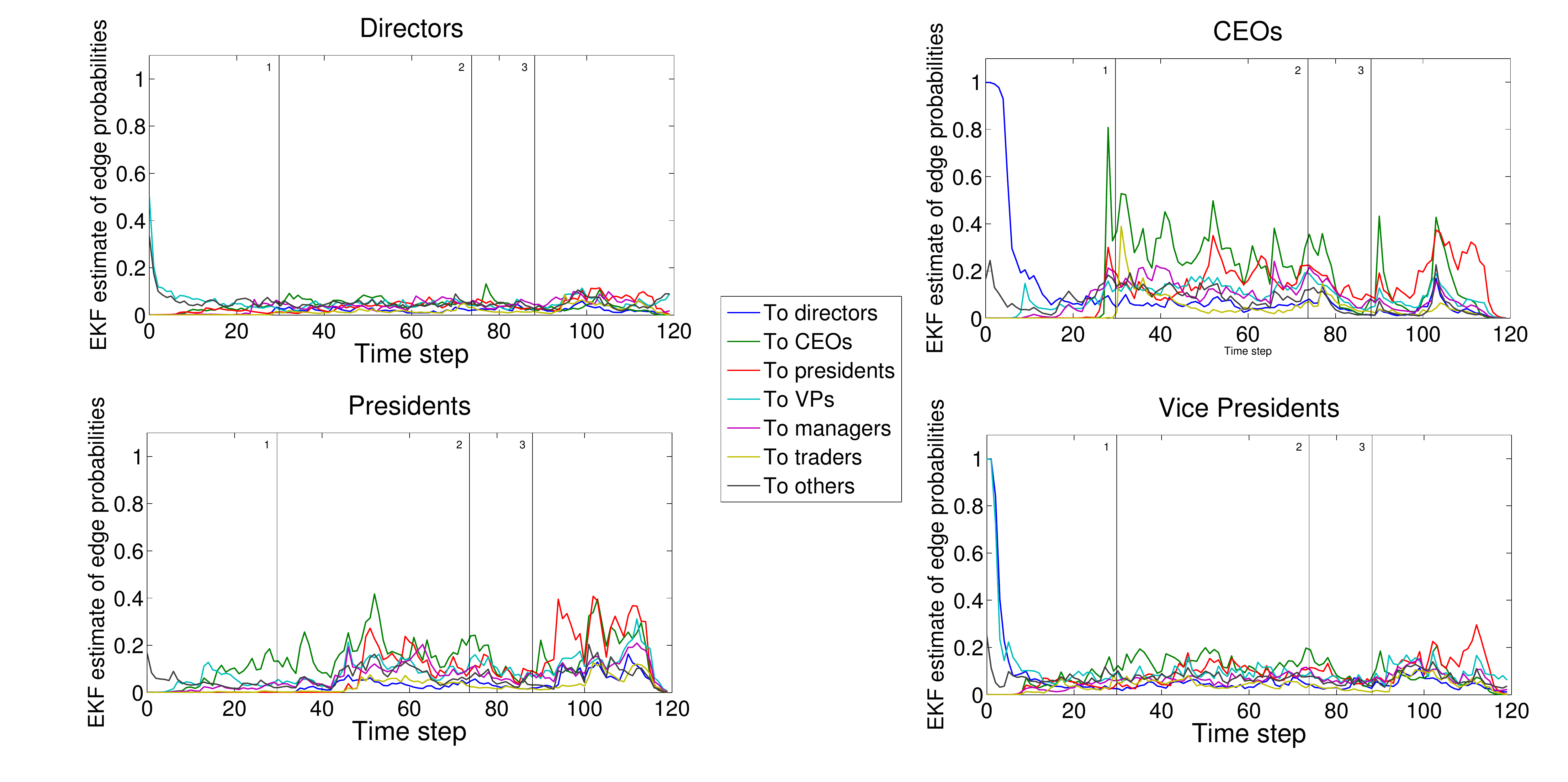}
\caption{Combined DSBM Results. These graphs show the results of combining the two layers of the network with a parameter $\alpha = 0.5$. Therefore, we should see attributes from both the behavioral and relational DSBM, and maybe some new, interesting results that result from combining the two layers.}
\label{fig_sim21}
\end{figure*}

For the relational DSBM parameters, the most interesting results come from the CEO's activity. Note that the CEO group combines all past and present CEO's. This evolution of parameters seems to indicate that during some of the important milestones in ENRON's demise, the CEO's were talking to each other more often, as well as sending out emails to the other employees in the network. This suggests that they were at least somewhat aware of what was happening with the company during these events, and had maybe discussed matters among themselves. From the relational layer, it also appears that the CEO's were the most active in communicating with other groups, where as the Directors showed very little connectivity. One explanation for this is that because the subset of employees that were studied were higher up in the company, the Director group didn't communicate with them as much, instead managing the lower level employees. Another interesting result is that the President group had much more activity towards the end of the time period, suggesting that as the legal situation worsened, their activity increased.

The behavioral DSBM parameters appear to be more noisy than their relational counterparts. In addition, they show very different behavior than the relational layer. The Vice Presidents appear much more active during the entire period when compared with the relational layer. Because of the nature of the TF-IDF and thresholding process, there could be a number of reasons for this. One possible reason could be that the weeks in which the Vice Presidents were active, they could have been sending a lot of forwarded emails, acting as a conduit of information between parties. This would cause the TF-IDF scores for the Vice President group to rise. 

Another interesting phenomenon in the behavioral layer is that of the CEOs. Specifically, it is interesting how their activity drops off significantly, and in fact one event that is very much apparent in the relational parameters completely disappears. This can only happen if the document content for the CEOs during those weeks are completely orthogonal to the other groups. Because we consider only text that the sender has written, and we only consider sent emails, one explanation could be that the CEO's forwarded many emails without adding any additional text. This would cause the list of words for the CEO to become very small.  However, a more likely explanation after some examination of the dataset shows that there is a large amount of activity in the relational dataset because many of the employees were emailing the CEO in a petition-like fashion, creating much activity. However, the CEO group actually sent very few emails during that time.

Combining the two networks as in Section~\ref{sec:posterior}, we run the DSBM for different levels of the mixing parameter $\alpha$. This was the probability of the selection variable choosing $W_1$ over $W_2$. Because of the use of binary networks in this example, the $\alpha$ parameter is used as the probability that the combined data will choose to use the relational network when the two layers disagree with each other. The objective in this particular example is to show that using this method we can not only reduce noise, but also discover interesting multifaceted behavior that is not obvious from one layer alone. We expect that this form of combination will emphasize traits or attributes that occur in both networks; however, attributes that exist mostly in one network but are strong enough will also be retained. We can study these effects through various network measures; in this case we look at betweenness and degree centrality.

Figure~\ref{fig_sim21} shows the DSBM parameters for mixing parameters $\alpha = 0.5$. Smaller values of $\alpha$ should be chosen because the relational network seems to be less noisy and more stable. This makes sense as the extrinsic relational interactions are directly measured. One interesting phenomenon that occurs is that much of the behavior that we saw in the relational layer is present, including the high level of CEO activity. However, the period of inactivity that is experienced in the behavioral layer for the CEO group has an effect by dampening the some of the strong peaks that we saw towards the end of the time period.

\begin{figure}[h!t]
\centering
\includegraphics[width = 2.9in]{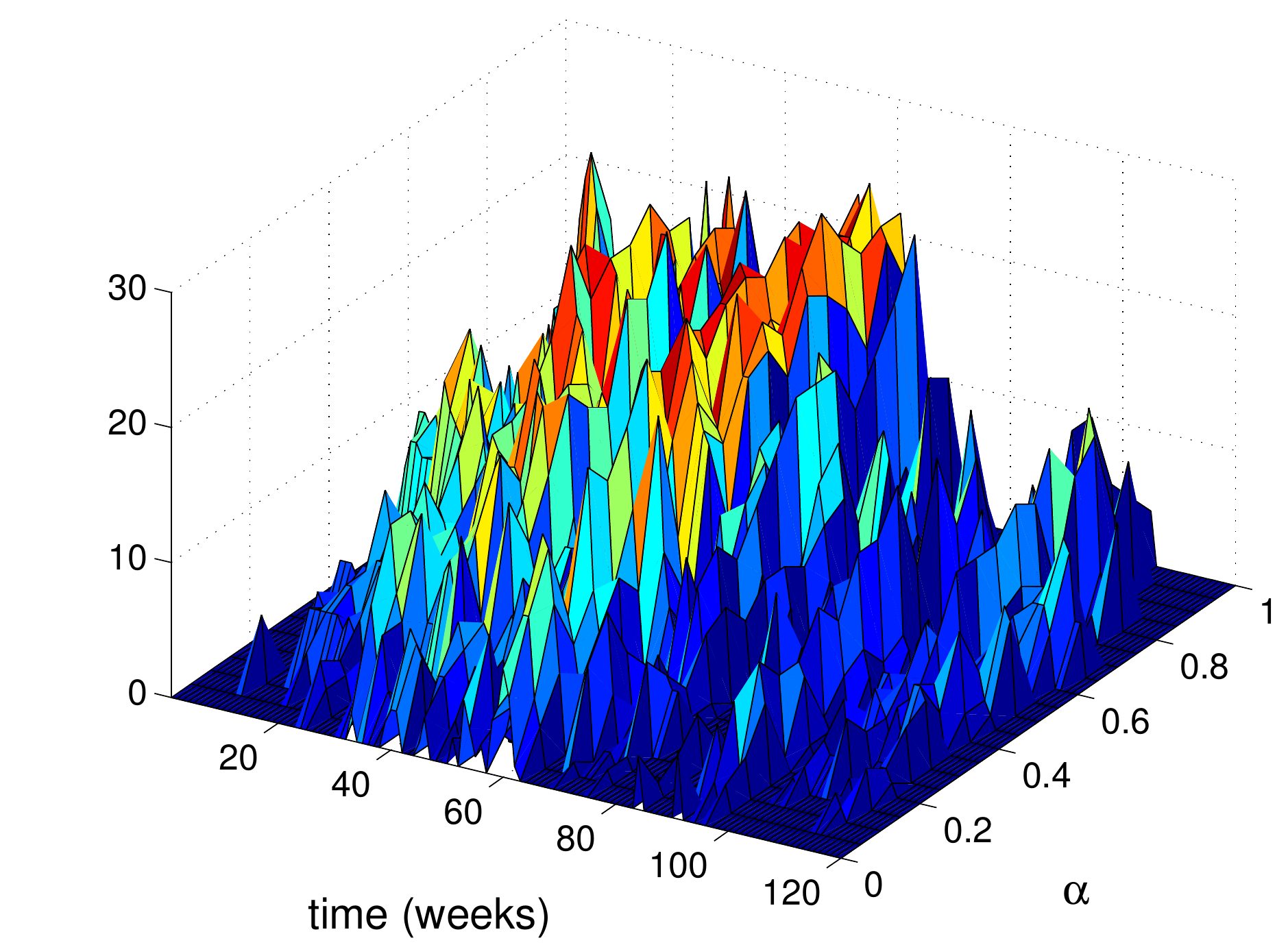}
\caption{Betweenness Centrality for Directors. This centrality is a measure of how connected a node is to the rest of the network. Larger centrality scores often occur for intermediate values of $\alpha$, particularly between time 95 and 115.}
\label{between}
\end{figure}

Figure~\ref{between} shows the betweeness centrality of the Directors group over time as the mixing parameter is varied. In general, the betweeness rises roughly monotonically as $\alpha$ is varied; however, from week 95 to week 115, betweenness centrality is significantly increased when using a combined dynamic network---that is, an intermediate value of $\alpha$. This time corresponds to the beginning of the company's upheaval and public disclosure of troubles. It may be concluded that by examining both network layers simultaneously we have removed some of the edges between other classes, and thus the centrality score of this particular group increased.  It is true that during this time, when overall email usage increased, the betweenness centrality measure went down, as there were more shortest paths through users from other groups. Using the combination of layers, however, there appears to be an increase in the number of shortest paths through the Directors group.

\begin{figure}[h!t]
\centering
\includegraphics[width = 2.9in]{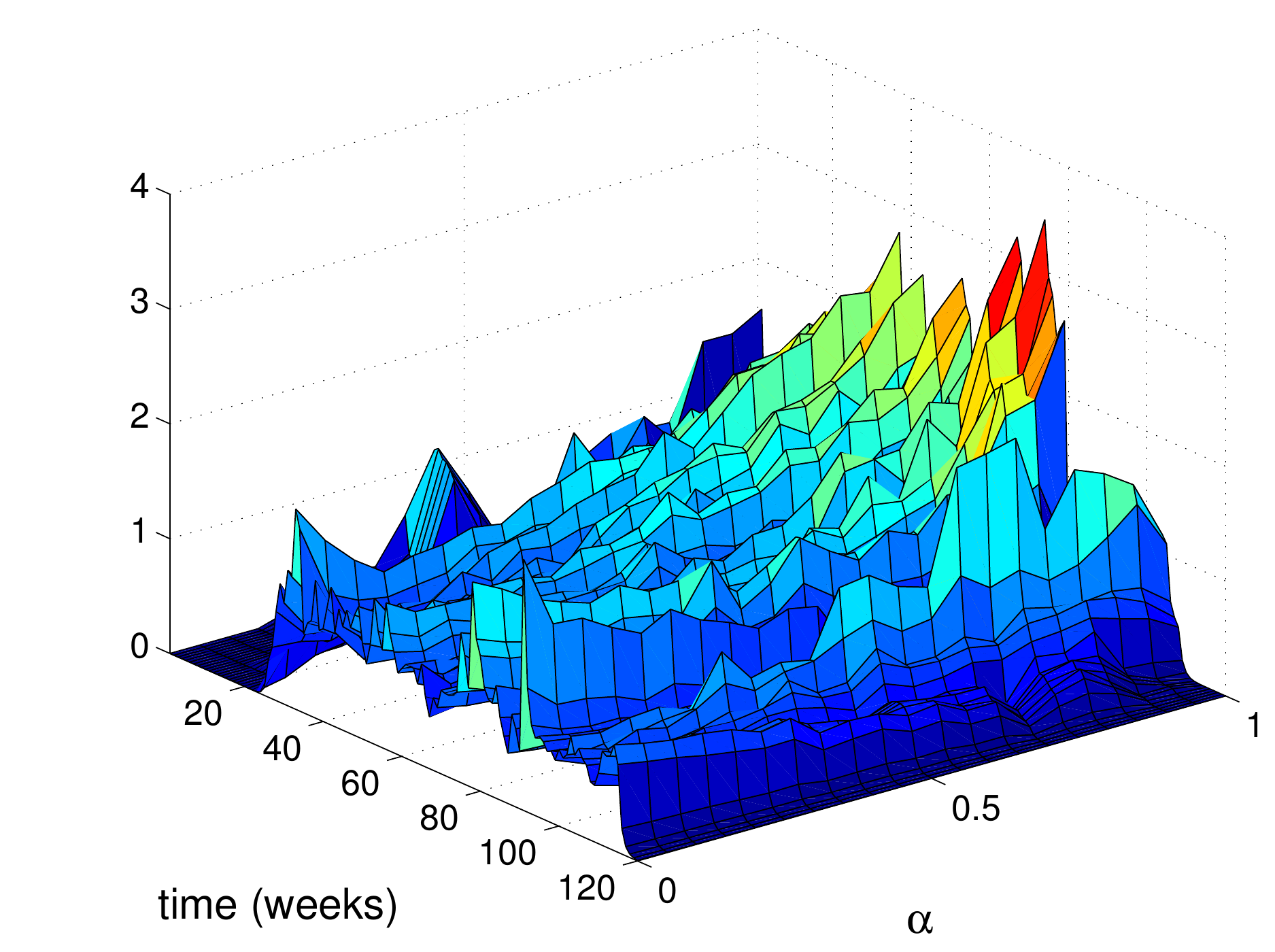}
\caption{Degree Centrality for CEOs. Higher degree centrality for $\alpha$ near one signifies greater activity in the behavioral network. Anomalous behavior can be seen in the later time steps as activity patterns shift.}
\label{degree}
\end{figure}

On the other hand, we can also see well-behaved monotonic behavioral correlations in some cases. Figure~\ref{degree} shows a transition of degree centrality for the class of CEOs (of which there were four during this time period). The behavioral network shows more connectivity for the CEO class. This phenomenon makes sense, as the behavioral data takes into account all written documents, which could be correlated with those of other users, while the relational network only takes into account direct communication between the CEOs and others.  In reality, much of that communication is performed through third parties (such as assistants), and thus CEOs probably do not send as much email as the average employee. Increasingly anomalous behavior occurs toward the end of the time period. We hypothesize that this is due to a larger volume of unusual emails sent directly to the CEO during this tumultuous period. 

\section{Related work}
The literature on single layer networks is large, with contributions coming from many different fields. There are many results on structural and spectral properties of a single-layer network, including community detection \cite{Ne04}, random walk return times \cite{NoRi04}, and percolation theory results \cite{AlBa02}. Diffusion or infection models have also been studied in the context of complex networks (see \cite{GuHaFa13}, for instance). 

Estimation of community structure in a network of agents is an active area of research in its own right. Specifically, the stochastic block model (SBM) \cite{HoLaLe83, WaWo87} is used to model community structure within a network by assuming identical statistical behavior for disjoint subsets of nodes. These communities are more flexible than simple cliques because it is not required that they be heavily interconnected, but only that they interact with nodes in other subcommunities uniformly. More recently, the SBM has been extended to track temporal changes in the network, appropriately called a Dynamic SBM, or DSBM. We follow the development in \cite{XuHe13}, but there have been other extensions of the classic SBM. In particular, \cite{YaChZh11} uses Gibbs sampling and probabilistic simulated annealing to estimate the Bernoulli parameters and class memberships over time. \cite{HoSoXi11} also fits a DSBM, but with a mixed membership model for the agents. The DSBM in \cite{XuHe13} uses an extended Kalman filter to track temporal changes between nodes, which will result in a smoothed and potentially insightful evolution of the estimated parameters.

Recently, there has been a growing interest in the multi-level network problem. Some basic network properties have been extended to the multilevel structure \cite{BaNiLa13}, \cite{Bi13} as well as some results that serve as an extension of single layer concepts, such as multi-level network growth \cite{NiBiLa13} and spreading of epidemics \cite{SaSeBo12}. The metrics that have been proposed attempt to incorporate the dependence of the layers into the statistical framework, which allows for a much richer view of the network. In the same vein, the approach described in this paper performs parameter inference on a multi-level network, incorporating some of the dependence information that the multi-level structure allows.

Bayesian model averaging is also related to this work; ideas from BMA are used to create conditional independence between the layers of a network \cite{raftery1995bayesian}. This framework accounts for the interdependent relationships between the multiple layers into latent variables, which can then be estimated.

\section{Conclusion}

We introduced a novel method for inference on multilayer networks. A hierarchical model was used to jointly describe the noisy observation matrices and MAP estimation was performed on the relevant latent variable. A simulation example using clustering demonstrated that the mixture of layers under the correct circumstances can lead to better results, and possibly a better understanding of the underlying structure between users. A real-life example was also discussed using the ENRON email dataset. The approach developed here can be extended to non-linear multi-objective optimization techniques to explore other ways of inferring multi-layer networks, such as Pareto ranking \cite{OsKuHe14} or posterior Pareto ranking \cite{HeFl08}.

\section{Acknowledgements}
We would like to thank Kevin Xu for providing the code for the DSBM model and his suggestions for utilizing it, as well as his general comments on the content of the paper.

\section{Appendix A: Solution of two gaussian distributions}

Theorem 1: Let $W \in \mathbb{R}^n$ The solution to the maximization problem

\begin{align}
	  \hat{W} = \mathrm{argmax}_W f(W) = \left[\gamma_1 P_1(W_1|W) + \gamma_2 P_2(W_2|W)\right]~,
\end{align}

with $P(W_i|W)$ of the multivariate Normal distribution

\begin{align}
  P(W_1|W)& = \mathcal{N}(W, \sigma_1^2 I_n) \\
  P(W_2|W)& = \mathcal{N}(W, \sigma_2^2 I_n)~,
\end{align}

is of the form 

\begin{equation}
\hat{W} = \beta W_1 + (1 - \beta)W_2,~~beta \in [0,1].
\end{equation}

Proof: The proof is separated into two steps. First, we show that for any arbitrary point $W \in \mathbb{R}^n$, the point $W_{\parallel}$ which is the projection of $W$ onto the line $g(W) = W_1 + \beta(W - W_1)$ increases the value of $f$, that is 

\begin{align}
  f(W) \le f(W_{\parallel})~.
\end{align}
Then we show that for all points on the line $g(W)$, $f$ is maximized for some point on the line segment between $W_1$ and $W_2$, corresponding to $\beta \in [0,1]$

Let $W \in \mathbb{R}^n$. There exists a unique decomposition of $x$ into a  vector parallel to $g(W)$ and one perpendicular to $g(W)$:

\begin{align}
W = W_{\parallel} +W_{\perp}~.
\end{align}

Plugging $W$ into $f(W)$, we have 

\begin{align}
f(W) &= \left(2\pi \right) ^{-n/2} | \sigma_1^2 I_n|^{-\frac{1}{2}}e^{-\frac{1}{2}(W - W_1)^T  (\sigma_1^2 I_n)^{-1}(W - W_1)} \nonumber\\
&\quad\quad + \left(2\pi \right) ^{-n/2} | \sigma_2^2 I_n|^{-\frac{1}{2}}e^{-\frac{1}{2}(W - W_2)^T  (\sigma_2^2 I_n)^{-1}(W - W_2)} \\
&=\left(2\pi \sigma_1^2\right)^{-n/2} e^{\frac{1}{2\sigma_1}  \|W_{\parallel} + W_{\perp} - W_1 \|^2} \nonumber \\
&\quad\quad + \left(2\pi \sigma_2^2\right)^{-n/2} e^{\frac{1}{2\sigma_2} \| W_{\parallel} + W_{\perp} - W_2\|^2}~.
\end{align}

The exponent can be decomposed as follows:

\begin{align}
(W_{\parallel} + &W_{\perp} - W_1)^T(W_{\parallel} + W_{\perp} - W_1)  \\
= (W_{\parallel}& - W_1)^T(W_{\parallel} - W_1) \notag\\
&+ 2W_{\perp}(W_{\parallel} - W_1) + x_{\perp}^T W_{\perp} \\
= (W_{\parallel}& - W_1)^T(W_{\parallel} - W_1) + W_{\perp}^T W_{\perp} \\
\ge (W_{\parallel}& - W_1)^T(W_{\parallel} - W_1)~.
\end{align}

Note that since $W_1$ is on the line $g(W)$, and $W_{\perp}$ is orthogonal to all points on $g(W)$, $W_{\perp}^TW_1 = 0$ and so the cross term goes to 0. The same can be shown for the other exponential term with $W_2$. Since the term with $x$ is greater than with just $x_{\parallel}$, so 

\begin{align}
f(W_\parallel) \ge f(W)~.
\end{align}

Finally, let us show that the maximum for $f$ must be between $W_1$ and $W_2$. This can easily be seen by the fact that both summation terms in $f$ decrease as the distance between $W$ and the means $W_1$ and $W_2$ increases. When on the line $g$, but outside the line segment between $W_1$ and $W_2$, moving closer to the means will increase both terms. Therefore, the maximum of $f$ must be on the line $g$, with $\beta$ restricted between 0 and 1.

\bibliographystyle{IEEEtran}
\bibliography{lib,CAMSAPbib}
\end{document}